\documentclass[preprint,aps]{revtex4}

\usepackage[latin1]{inputenc}
\usepackage{graphicx}

\begin{document}
\title{Bloch vector dependence of the plasma frequency in metallic photonic crystals}
\author{Didier Felbacq}
\affiliation{Groupe d'\'Etude des Semiconducteurs UMR-CNRS 5650 \\
Bât. 21 CC074, Place E. Bataillon 34095 Montpellier Cedex 05, France}

\author{Fr\'ed\'eric Zolla}
\affiliation{Institut Fresnel \\
UMR 6133, Case 162, Universit\'e Aix-Marseille I, Av. Escadrille
Normandie Niemen}

\author{Guy Bouchitt\'e}
\affiliation{Laboratoire ANLA \\
 Université de Toulon et du Var \\
La Garde Cedex, France}
\date{\today }

\begin{abstract}
We show that the plasma frequency in wire photonic crystals depends
upon the Bloch vector. An accurate formula is given.
\end{abstract}
\maketitle
\section{Introduction}
Photonic crystals are artificial materials structured periodically, with the aim
at controlling light propagation in them \cite{dowling}.
Metallic photonic crystals \cite{yablo} have recently attracted the attention of several
authors for their ability of exhibiting a very low plasma frequency and a
homogeneous behavior corresponding to a negative relative permittivity \cite
{mcphed,mcphed2,mcphed3,pendry,pendry2,moroz,moi1,moi2,moi3,moi4,soukou,pokro}. More recently, indications of a
negative permeability have been reported \cite{smithtera,obrien,smith,val,muneg,felbou}. 
The mixture of both media lead to
left-handed materials \cite{veselago}. In a recent paper \cite{soukou}, Soukoulis et al. have inspected
various formulae for deriving the homogenized permittivity and the cut
frequency for a system of parallel metallic wires. Various approaches have
been proposed, all of them leading to different formulas, none of which being
really fully satisfactory. In this letter we provide an accurate
formula for the cut-frequency as a function
of the angle of incidence (or, more generally, the horizontal component of the Bloch vector), for a wire mesh
 photonic crystal. 
The wire mesh photonic crystal is modelized as a stack of diffraction
gratings made of very thin rods (radius $a$) with period $d$ (see fig. 1). It is illuminated
by an incident, s-polarized, plane wave $E^{i}$ under the incidence $\theta$. The wavelength
in vacuum is $\lambda$ and the wavenumber is $k_{0}=2\pi/\lambda$. We denote by
$E^t$ the total electric field and by $E^d=E^t-E^{i}$ the diffracted field.
The basic assumption here is that $k_{0}a \ll 1$ and $k_{0}d \ll 1$.

\section{Asymptotic analysis of the scattered field}

The field diffracted at ${\bf r}=(x,y)$ by the wire situated at abscissa $n \times d$ reads as:
$E_{n}^{d}({\bf r})=b_{n}H_{0}^{(1)}(k_{0}|{\bf r}-nd {\bf e}_{x}|)$. 
Here the wires are periodically settled and thanks to Bloch
theorem we have: $b_{n}=e^{ink_{0}d\sin \theta } b_{0}$, where $b_{0}$ is a coefficient to be 
determined. Adding the contribution of all wires, we get, for the total diffracted field:
\begin{equation}
E^{d}({\bf r})=b_{0}\sum_{n} e^{ink_{0}d\sin \theta}H_{0}^{(1)}(k_{0}|{\bf r}-nd {\bf e}_{x}|) 
\end{equation}
From scattering theory \cite{feljos}, we know that the field diffracted by one rod
is obtained through the scattering matrix $s_{0}(k_{0}a)=-J_{0}(k_{0}a)/H_{0}^{(1)}(k_{0}a)$ by:
$\, b_{n}=s_{0}[E^{i}(nd {\bf e}_{x})+\sum_{m\neq n} b_{m} H_{0}^{1}(k_{0}|m-n|d)]$, that is:
\begin{equation}\label{bo}
b_{0}=\frac{ E^{i}(0)}{s_{0}^{-1}-\sum_{n \neq 0} e^{ink_{0}d\sin \theta}H_{0}^{(1)}(k_{0}|n|d)}
\end{equation}

($J_0$ and $H_0^{(1)}$ are respectively the Bessel function of $0^{th}$
order and the Hankel function of first type and $0^{th}$ order).

The total diffracted field is given by: 
\begin{eqnarray}\label{total}
E^d\left( {\bf r}\right) =b_0\sum_n e^{i nk_{0}d\sin \theta} 
H_0^{( 1)}\left( k_0 \left| {\bf r}-nd{\bf x}\right| \right) 
\end{eqnarray}

Our aim is now to study the limit of this expression when the wavelength is
large with respect to the radius of the rods and the period of the grating.
The fundamental quantities are then $k_0 a$ and $k_0 d$. For later
purpose, we define the following combined quantities: 
\begin{eqnarray}
 L=
\frac{d}{\pi }\ln \left ( \frac{d}{2\pi a} \right )\, , \,
\chi = \beta_0 L
\end{eqnarray}
where $\beta_0=k_0 \cos\theta$.

The point is to evaluate the asymptotic behavior of $b_0$ and that of the
series defining $E^d ( {\bf r} ) $. For that purpose, we note the
following representation formula: 
\begin{equation}\label{poisse}
\sum_n e^{i n \alpha_{0} d} H_0^{( 1) }
\left( k_0 \left| {\bf r}-n d{\bf e}_{x}\right| \right)
 =\frac{2}{d}\sum_n \frac{1}{\beta_n}
e^{i(\alpha_n x+\beta_n \left| y\right|) }
\end{equation} 
where $\beta_n=\sqrt{k_0^2-\alpha_n^2},$ 
$\alpha_n=k_0\sin\theta +n\frac{2\pi }{d}$.

The behavior of $b_0$ when $k_0 d$ tends to $0$ cannot be obtained by
taking the limit of each term of the series in (\ref{bo}), because the terms are singular there.
 Instead, we perform an asymptotic analysis, 
starting with equality (\ref{poisse}). Our point is to let ${\bf r}=(x,y)$ tend to $0$,
and we first set $y=0$:
\begin{equation}\label{poiss2}
\sum_{n\neq 0} e^{i n \alpha_{0} d} H_0^{( 1) }
\left( k_0 \left| x-n d\right| \right)
 =-H_{0}^{(1)}(k_{0}|x|)+\frac{2}{d}\sum_n \frac{1}{\beta_n}e^{i\alpha_n x}
\end{equation} 
We cannot let $x$ tend to $0$ in the right hand side, because $H_{0}^{(1)}(k_{0}|x|)$ and the series
are singular there. However, both singularities compensate.
Indeed, the singularity of the series can be obtained explicitely by analysing 
the convergence ratio of the series: as $n$ tends to infinity we have $\beta_{n} \sim 2i\pi |n|/d$,
so the series is logarithmic:
$$\frac{2}{d}\sum_n \frac{1}{\beta_n}e^{i\alpha_n x}=
\frac{2}{d\beta_0}e^{i\alpha_0 x}+
\frac{2}{d}\sum_{n\neq 0} \left(\frac{1}{\beta_n}-\frac{d}{2i\pi |n|}\right)e^{i\alpha_n x}+
\sum_{n\neq 0} \frac{1}{i\pi |n|}e^{i\alpha_n x}
$$
 and the last series is easily seen to be equal to :
 $\frac{2}{i\pi}e^{i\alpha_{0} x}\ln [2 \sin (\pi x/d)]$.
By using the expansion of $H_{0}^{(1)}(k_{0}|x|)$ near $x=0$, we get ($\gamma$ is the Euler constant):
\begin{equation}
\begin{array}{ll}
-H_{0}^{(1)}(k_{0}|x|)+e^{i \alpha_{0} x }\frac{2}{i\pi}\ln[2 \sin (\pi x/d)]
& \sim
-1-\frac{2i}{\pi} \gamma+\frac{2i}{\pi} \ln(\frac{2\lambda}{d})
\end{array}
\end{equation}
which shows that : 
\begin{equation}\label{exp}
\sum_{n \neq 0}e^{i n \alpha_{0} d}H_0^{( 1) }
\left( k_0\left|n\right| d\right) =
-1-\frac{2i}{\pi} \gamma+\frac{2i}{\pi} \ln(\frac{2\lambda}{d})
+\frac{2}{d\beta_0}+
\frac{2}{d}
\sum_{n> 0} \left(\frac{1}{\beta_n}+\frac{1}{\beta_{-n}}-\frac{d}{i\pi |n|}\right)
\end{equation}

Finally, when $k_{0}d$ is small, the last series is equivalent to 
$$\frac{i [-3+2\cos^{2}(\theta)]}{\pi^3}\zeta(3) (k_{0}d)^2$$
and consequently
\begin{equation}\label{exp2}
\sum_{n \neq 0}e^{i n \alpha_{0} d}H_0^{( 1) }
\left( k_0\left|n\right| d\right) =
-1-\frac{2i}{\pi} \gamma+\frac{2i}{\pi} \ln(\frac{2 \pi}{k_0 d})
+\frac{2}{d\beta_0}+
\mathcal{O}((k_{0}d)^2)
\end{equation}

We are now in a position to give an asymptotic expansion for $b_{0}$.
When $k_0 a$ tends to zero, we have: 
\begin{equation}
s_0 ^{-1}( k_0 a ) 
= - \frac{H_0^{(1)}(k_0 a)}{J_0(k_0 a)}=
-1-\frac{2i}{\pi }(\gamma+\ln ( k_0 a /2)) + \frac{i}{2 \pi} (k_0a)^2
+ \mathcal{O}((k_0 a)^3)  \; .\label{szero}
\end{equation}
From this expression and (\ref{exp2}), where we remove the terms that tend to $0$  with $k_{0}d$, we obtain
from (\ref{bo}):
\begin{equation}\label{boeq}
b_0 \sim -\frac{\beta_0 d}{2}
\frac{1}{1 -i\chi }E^i ( 0)
\end{equation}

We are now able to give an asymptotic expansion of $E^d$ by splitting
the sum (\ref{total}) into two terms corresponding to the evanescent modes and the
unique propagative mode :
\begin{equation}
  \label{eq:Edsplit}
  E^d \sim E^d_{prop} + E^d_{evan}
\end{equation}
where
\begin{equation}
  \label{eq:Edprop}
  E^d_{prop} = \frac{-1}{1-i\chi}\, e^{i(\alpha_0 x + \beta_0 |y|)} E^i(0) \, , \,
   E^d_{evan} = -\frac{2 b_0}{d}
\sum_{n \ne 0} \frac{1}{\beta_n} e^{i(\alpha_n x + \beta_n |y|)}
E^i(0) \, .
\end{equation}
The total field is obtained by adding the incident field:
\begin{equation}
\begin{array}{cl}
\hbox{For } y>0: &  E^{t}({\bf r})= \left [ e^{i(\alpha_{0}x-\beta_{0}y)}+
\frac{-1}{1-i\chi}e^{i(\alpha_{0}x+\beta_{0}y)}
-\frac{1}{L} \frac{\chi}{1-i \chi} \sum_{n \ne
  0}\frac{1}{\beta_{n}}e^{i(\alpha_{n}x+\beta_{n}|y|)} \right ] E^i(0) \\
\hbox{For } y<0: & E^{t}({\bf r})= \left [ \frac{\chi}{\chi+i}e^{i(\alpha_{0}x-\beta_{0}y)}
-\frac{1}{L} \frac{\chi}{1-i \chi} \sum_{n \ne
  0}\frac{1}{\beta_{n}}e^{i(\alpha_{n}x+\beta_{n}|y|)} \right ] E^i(0)
\; .
\end{array}
\end{equation}
The reflection and transmission coefficients are readily obtained:
\begin{equation}\label{rt}
r=\frac{-1}{1-i\chi} \, , \, t=1+r=\frac{\chi}{\chi+i} 
\end{equation}

The expression (\ref{rt}) shows that the  reflection
coefficient tends to $-1$ as  $k_{0}d$ tends to  $0$. This result is
quite a striking one if one thinks of the extremely low concentration of
material in this scattering experiment.   The behavior of such a grid
is equivalent to a  perfect mirror at the vicinity of $\lambda=+\infty$. For
instance, for $a/d=1/1000$ and for $k_0 d=1/100$, in normal
incidence we find a theoretical reflection coefficient worthy of the
best mirrors: $R=|r|^2=0.999739$ !  
Moreover the expression of the reflection coefficient $r$ gives us
the critical dimension of the radii of the wires. 
If $a(k_0)$ is related to $k_0$ in such a way that 
\begin{equation}
  \label{eq:critic}
\frac{k_0 d \cos \theta_0}{\pi} \log \bigl (
  \frac{d}{2 \pi a(k_0)} \bigr )=\Gamma  
\end{equation}
where $\Gamma$ is some constant, i.e. 
\begin{equation}
  \label{eq:acritic}
 a(k_0)= \frac{d}{2 \pi} e^{-\frac{\pi \Gamma}{k_0 d \cos \theta_0}}  
\end{equation}
then the grid, at the limit, does not behave neither as vacuum nor as a perfect
mirror because the reflection coefficient is equal to $r =\frac{-1}{1+i \Gamma}$. 
Coming back now to the evanescent field and
having in mind that $k_0 d \ll 1$ we have $\beta_n \sim 2i\pi |n| /d$ and
therefore the expression written above (\ref{eq:Edprop}) can still be simplified:

\begin{equation}
\label{Edevanfin}
E^d_{evan} \sim \frac{-4 i}{\log \left( \frac{d}{2 \pi a}\right )} 
\log \left ( 1- e^{iK (x + i |y)|} \right )
E^i(0)
\end{equation}

\section{Derivation of the cut-wavelength}
In the preceding  section, we have derived an  explicit expression for
the field diffracted by the wire grating, this allows us to derive the
dressed $T_G$ matrix \cite{felbou} of a single grating layer. This matrix
does not  depend upon the  characteristics of the incident  wave, that
is, the wavelength and the angle of incidence.
\begin{equation}\label{eq:TG}
T_G=\left(
\begin{array}{cc}
1 & 0 \\ 
\frac{2}{L} & 1
\end{array}
\right) 
\end{equation}
Finally, in order to modelize the wire mesh photonic crystal, we derive
the total dressed $T$ matrix by adding a slab of air
below the wire mesh grating. The $T_h$ matrix for a homogeneous slab of 
dielectric material with permittivity $1$ between $0$ and $-h$ is given by: 
\begin{equation}
T_h=
\left( 
\begin{array}{cc}
\cos ( \beta_0 h)  & \beta_0^{-1} \sin ( \beta_0 h )  \\ 
-\beta_0 \sin ( \beta_0 h )  & \cos ( \beta_0 h ) 
\end{array}
\right) 
\end{equation}
so that the total $T$-matrix for the grating and the slab is: 
\begin{equation}
T=\left( 
\begin{array}{cc}
\cos ( \beta_0 h) +\frac{2}{\beta_0 L}\sin ( \beta_0 h ) 
& \beta_0 ^{-1} \sin ( \beta_0  h )  \\ 
-\beta_0  \sin ( \beta_0 h) +\frac{2}{L}  
\sin ( \beta_0 h )  & \cos ( \beta_0 h ) 
\end{array}
\right) 
\end{equation}

We have now characterized a basic layer of the photonic crystal. 
A general device is made of a stack of $N$ such layers. 
Due to the very weak evanescent fields, the transfer matrix of 
such a layered device is very well approximated by $T^N$. Note that
the influence of the evanescent waves can be easily derived from Eq. \ref{Edevanfin}.

In wire mesh photonic crystals, it has been well establish that there is 
a cut-wavelength, above which there is a band gap (this phenomenon is sometimes
described as a plasmon frequency). Our point is to derive an implicit equation for the plasmon
wavelength $\lambda_{c}$. This wavelength corresponds to the edge of a band gap.
A band gap is characterized by the fact that
$\left| tr(T) \right|\ge 2$, when the wavelength is very large with respect to $d$,
the transfer matrix is very near the identity matrix, consequently, the equation for the edge $\lambda_{c}$ of the 
last gap reads as: $tr ( T(\lambda,\theta) ) =2$. This amounts to looking for
$\beta_0^c=\frac{2\pi}{\lambda_{c}}\cos \theta$ which is solution of:
\begin{eqnarray}
\cos ( \beta_0^c h) +\frac{1}{\beta_0^c L }\sin ( \beta_0^c h )=1.  
\end{eqnarray}
In order to solve this equation we denote:$ X=\tan( \beta_0^c h/2 ) $, 
and we obtain: 
\begin{equation}
\frac{1-X^2}{1+X^2}+\frac{1}{\chi }\frac{2X}{1+X^2}=1
\end{equation}
whose solution is $\beta_0^c L =X$. Finally, we get
that the plasmon frequency is given by the following
implicit dispersion relation: 
\begin{equation}\label{eq:disprel}
\beta_0^c L  \tan \left( \beta_0^c h/2 \right) =1 \; .
\end{equation}
Now, if we let $x^c=\frac{h \beta_0^c}{2 \pi}$, this dimensionless
number is solution of:
\begin{equation}\label{eq:xc}
2 \pi x^c \frac{L}{h} \tan \left ( \pi x^c\right )=1 \; ,
\end{equation}  
that we have to solve numerically for any parameter
$\frac{L}{h}$. Three cases can therefore occur.
\begin{itemize}
\item
$\frac{L}{h} \ll 1$ (\textit{i.e.} $h \gg d$, for realistic size of
rods, say $a=10^{-3} d$). In that case, we have $x^c \sim 1/2$ and, as
a result, the cut-wavelength $\lambda_0^c$ is given by:
\begin{equation}
\lambda_0^c= \frac{2 \pi \cos \theta}{\beta_0^c}  \sim 2 h \cos \theta
\; .
\end{equation}
which is nothing but the Bragg condition.
As a conclusion $\lambda_0^c \gg d$ (except for grazing incidence)
which is, fortunatly, compatible with the homogenization process.

\item
$\frac{L}{h} \gg 1$. In that case we can make equation \ref{eq:xc}
explicit by using the expansion $\tan( \pi x^c) \sim \pi x^c$. We
find:
\begin{equation}
x^c= \frac{1}{\sqrt{2} \pi} \sqrt{\frac{h}{L}}
\end{equation}  
and we deduce an approximation $\lambda_0^{c,1}$ of $\lambda_0^{c}$
:
\begin{equation}\label{eq:lambda0C1}
\lambda_0^{c,1}= \pi \sqrt{2 L h} \cos \theta
\end{equation} 

\item
$\frac{L}{h} \sim 1$. We are therefore between the two previous
cases. And in order to test the accuracy of our formula, we have to
make some numerical experiments which is the topic of the next
paragraph.
\end{itemize}

In this paragraph, our aim is to evaluate the accuracy of
Eq. \ref{eq:xc} in some representative examples. Take the example of
the stack of gratings each of them being made of thin circular
metallic rods.  In the following
$\frac{a}{d}=0.005$ has been taken for different ratios $\frac{h}{d}$
and different kind of crystals defined by the angle $\psi$  in order to check the validity of our
method. For this purpose, we plot the $0-$order efficiency versus the
normalized wavelength $\frac{\lambda}{d}$ in normal incidence as shown
in Figs. 1-3. It is of importance to note
that the cut-wavelength $\lambda_0^c$ derived from
Eq. \ref{eq:disprel} is extremely reliable even if $\frac{d}{h} > 1$
as shown in Fig. 1 for which 
$\frac{d}{h}=2$. Conversely, the Bragg condition leads to absurd results except
for very large ratios $\frac{h}{d}=2$. 

%\section{Impedance relation and its consequencies}
Neglecting evanescent waves as we did to obtain the
cut-wavelength, we can found an equivalent formulation \textit{as
  per}:
\begin{equation} \label{eq:jump}
\left [E^d \right ]_{y=0}=0, \quad \hbox{and} \quad
\left [\frac{\partial E^d}{\partial y} \right ]_{y=0}= \frac{2}{L} E^d(0) 
\end{equation}
where $\left [ \cdot \right ]_{y=0}$ is the jump of the quantity
within the brackets when crossing the plane $y=0$. We have, indeed,
$E^d$ solution of $\triangle E^d + k_0^2 E^d=0$ everywhere except for
$y=0$ and due to
the invariance of the above relations with respect to $x$, we can look
for solutions in the form $E^d(x,y)= e^{i\alpha_0 x} U(y)$. The
outgoing waves lead therefore to:
\begin{equation}
U(y)=
\left \lbrace
\begin{array}{cc}
1 + r_0 e^{i \beta_0 y} & , y>0 \\
t_0  e^{-i \beta_0 y} &, y<0
\end{array}
\right .
\end{equation}
Eventually, by making use of Eqs. \ref{eq:jump}, we find expressions
which are consistent with Eq. \ref{eq:TG}, namely:
\begin{equation}
r_0 = \frac{-1}{1 - i \beta_0 L} \quad \hbox{and} \quad
t_0 = 1 + r_0 \; .
\end{equation}
As a conclusion, $E^d$ is solution in $\mathcal{D}'$ (sense of
distributions) of:
\begin{equation}
\triangle E^d + k_0^2 E^d= \frac{2}{L} E^d \delta_{|y=0} \; ,
\end{equation}
and more generally for a stack of $N$ gratings
\begin{equation}
\triangle E^d + k_0^2 E^d= \frac{2}{L} E^d \sum_{k=0}^{N-1}
\delta_{|y=-k h} \; .
\end{equation}

\section{Concluding remarks}
In this paper, we derive a new expression for perfectly metallic
gratings at long wavelengths which leads to a very simple dressed
matrix associated with one grating. In a second step we find a very
accurate formula which gives the cut-wavelength of a very sparse
photonic crystal even for cut-wavelengths of the same magnitude of the
period of  aforementionned crystals. This method can be easily
generalized for non circular rods ; the characteristic length $L$
cannot be derived in a closed formula and therefore necessitates a
numerical computation. Moreover, the involved structures being
infinitly conducting and completely transparent for the
$p-$polarization, we think that it is possible to derive explicit
formulae for three-dimensional structures. 

\newpage
Figures captions\\

Figures captions\\

Figure 1: Grid composed of thin infinitly metallic circular rods.\\

Figure 2: $0-$order efficiency $R_0$ versus the wavelength
    $\lambda$ for a stack of 10 gratings in normal incidence for
    $d=1$, $a=0.005$ and $h=0.5$. $\lambda_0^{\mathrm{Bragg}}=2 h=1$
    represents the Bragg wavelenghth, whereas $\lambda_0^{c,1}$
    represents the first approximation given by Eq. \ref{eq:lambda0C1} and
    $\lambda_0^{c}$ is the cut-wavelength derived from Eq. \ref{eq:disprel}.\\
Figure 3:     $0-$order efficiency $R_0$ versus the wavelength
    $\lambda$ for a stack of 10 gratings in normal incidence for
    $d=1$, $a=0.005$ and $h=1$. $\lambda_0^{\mathrm{Bragg}}=2 h=2$
    represents the Bragg wavelenghth, whereas $\lambda_0^{c,1}$
    represents the first approximation given by Eq. \ref{eq:lambda0C1} and
    $\lambda_0^{c}$ is the cut-wavelength derived from Eq. \ref{eq:disprel}.\\
Figure 4:    $0-$order efficiency $R_0$ versus the wavelength
    $\lambda$ for a stack of 10 gratings in normal incidence for
    $d=1$, $a=0.005$ and $h=1$. $\lambda_0^{\mathrm{Bragg}}=2 h=10$
    represents the Bragg wavelenghth, whereas $\lambda_0^{c,1}$
    represents the first approximation given by Eq. \ref{eq:lambda0C1} and
    $\lambda_0^{c}$ is the cut-wavelength derived from
    Eq. \ref{eq:disprel}. Note the widths of gaps for this very sparse
    structure (Filling ratio=$1.57\, 10^{-5}$!)\\     
\newpage
\begin{figure}[h]
    \includegraphics*[width=6cm,height=6cm]{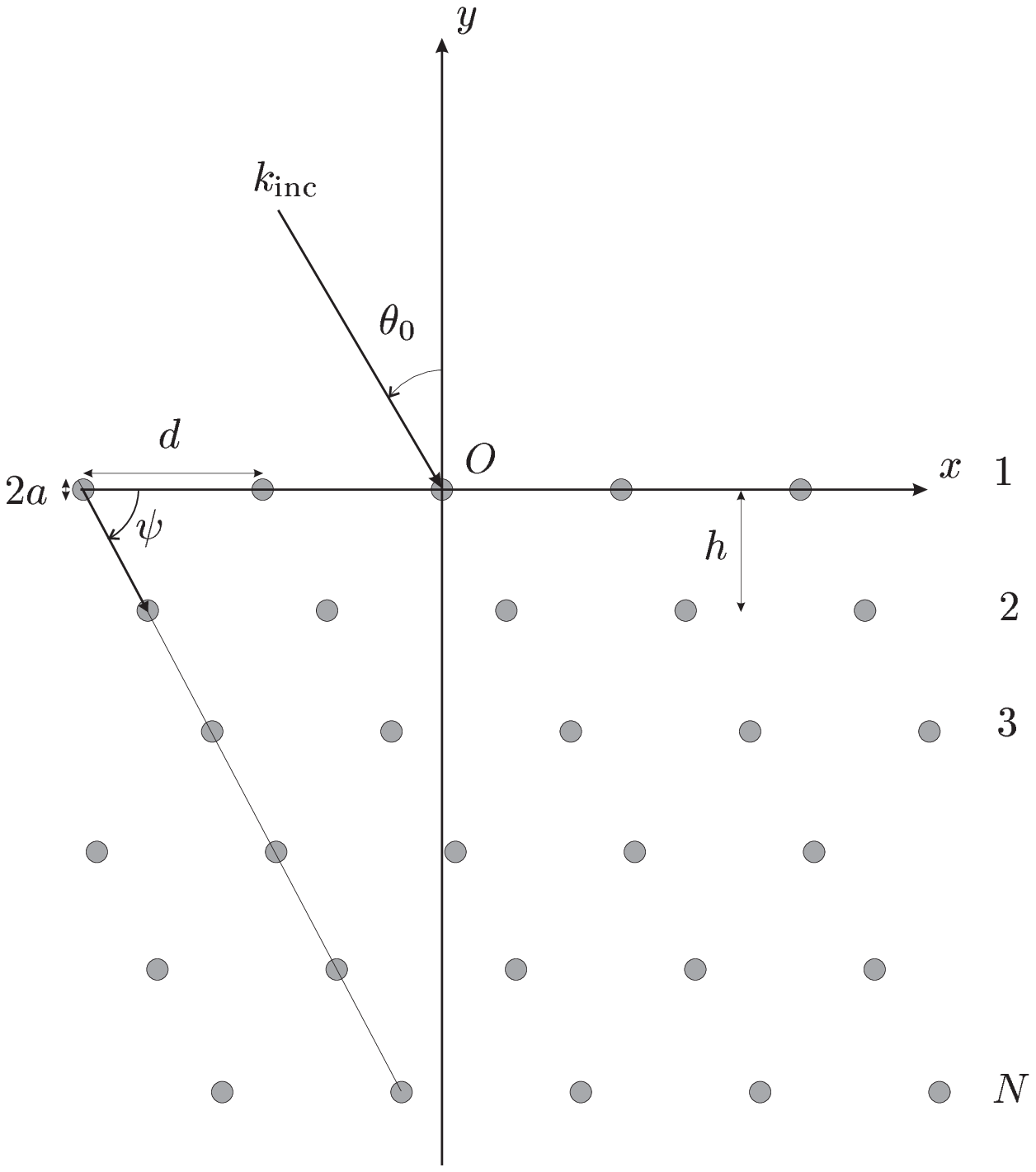}
 \end{figure}
 \begin{figure}[h]
    \includegraphics*[width=6cm,height=6cm]{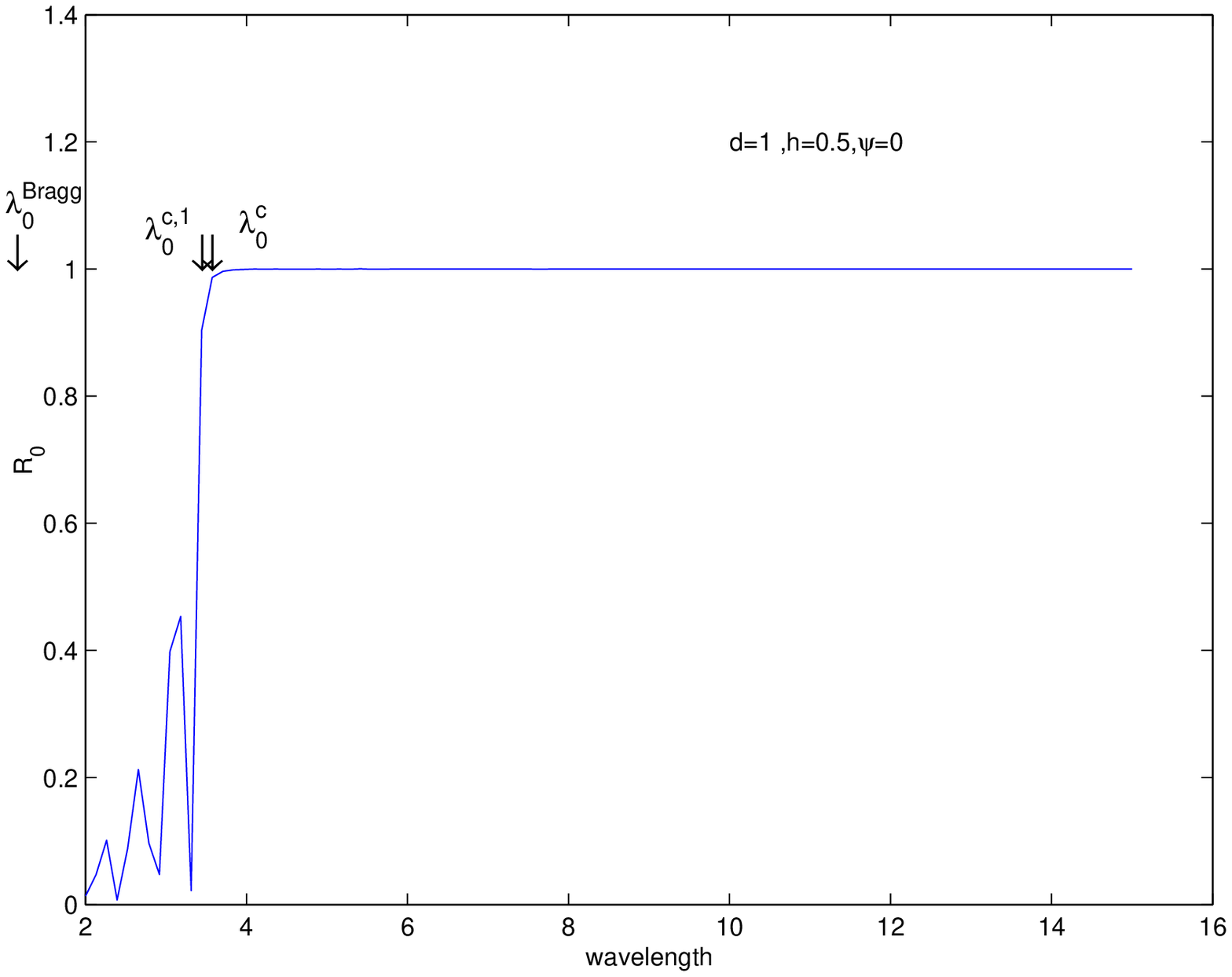}
   \end{figure}
 \begin{figure}[h]
    \includegraphics*[width=6cm,height=6cm]{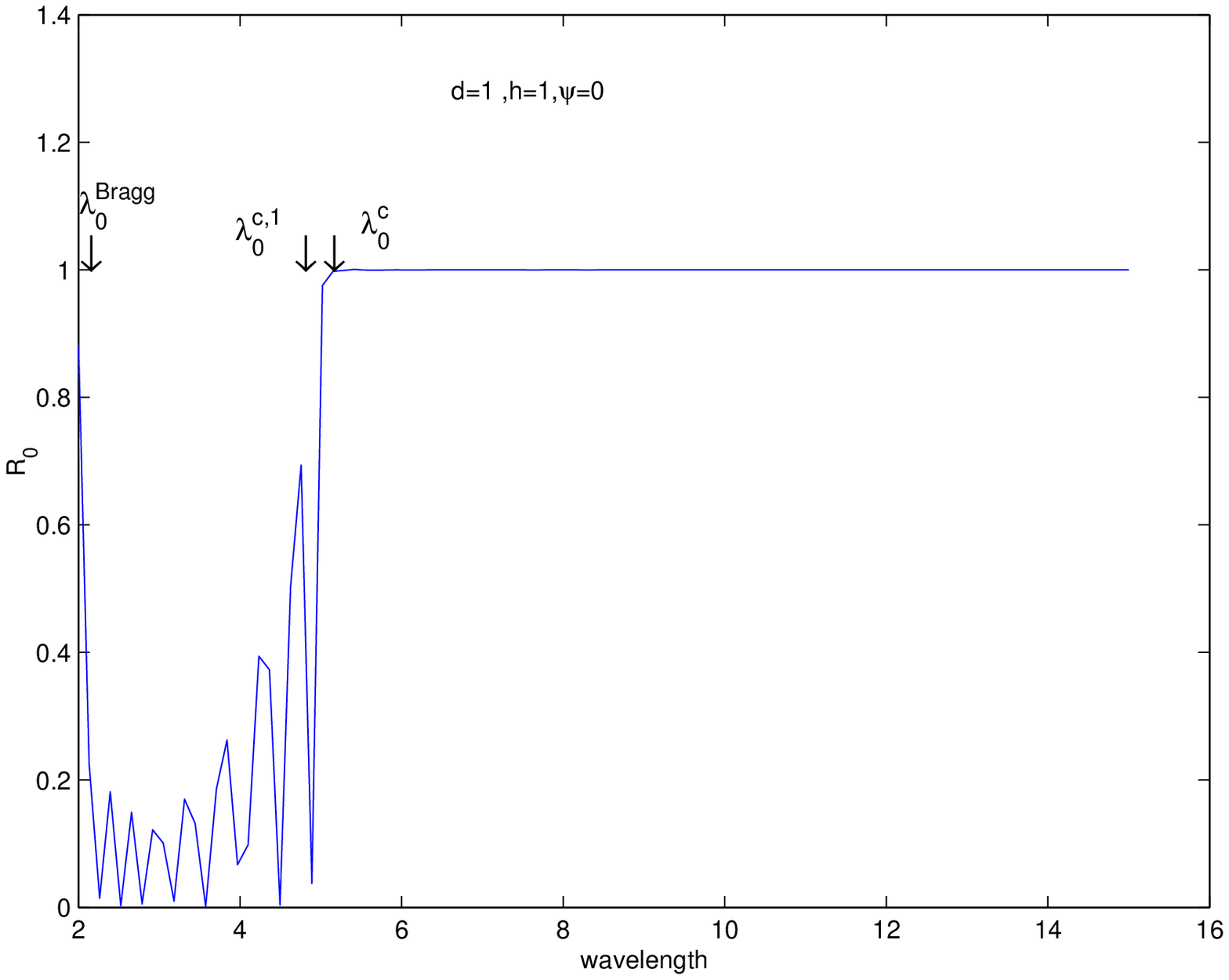}
     \end{figure}
 \begin{figure}[h]
    \includegraphics*[width=6cm,height=6cm]{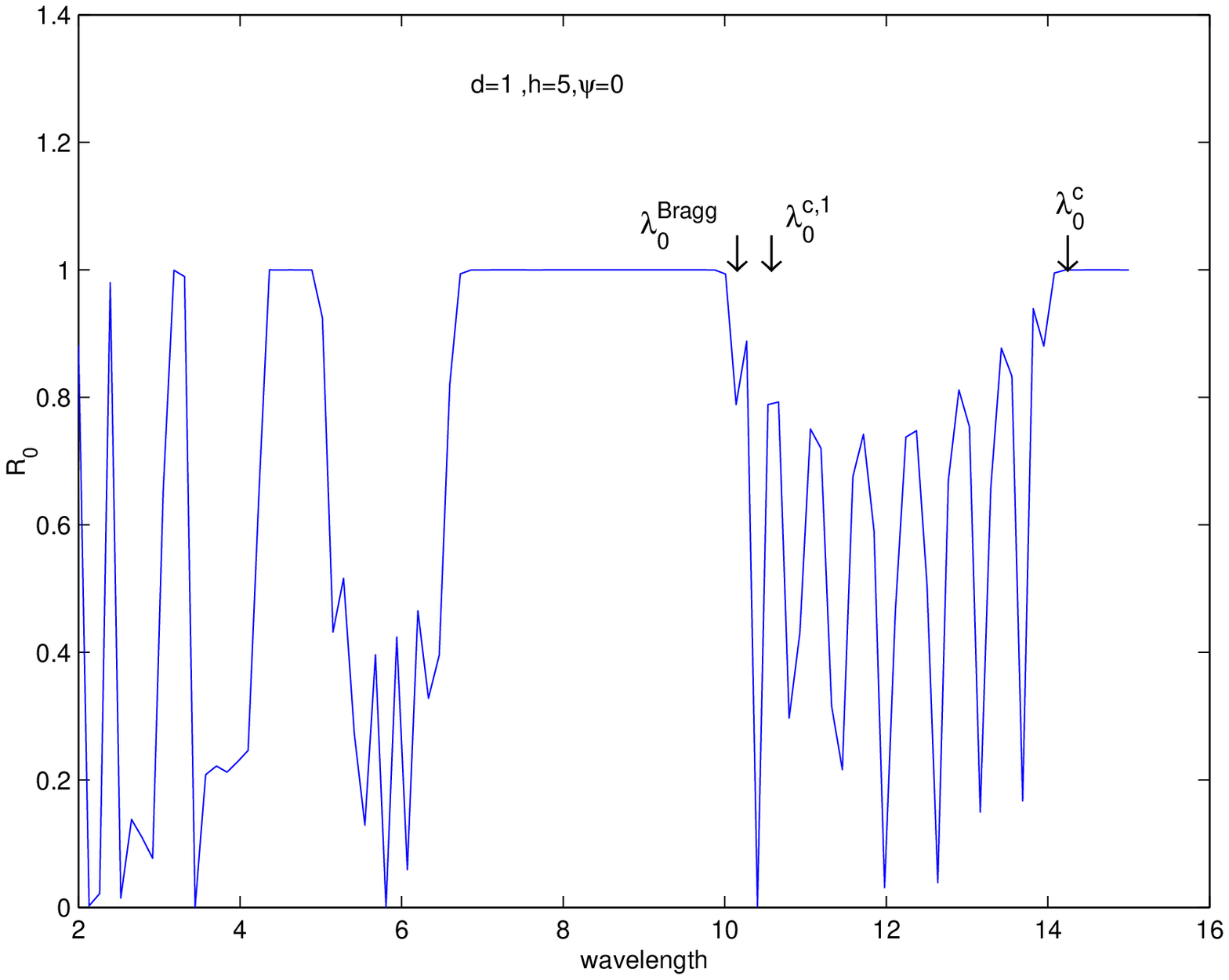}
     \end{figure}

\begin{thebibliography}{}
\bibitem{dowling} http://phys.lsu.edu/~jdowling/pbgbib.html
\bibitem{yablo} D. F. Sievenpiper, M. E. Sickmiller, and E. Yablonovitch, Phys. Rev. Lett. {\bf 76}, 2480 (1996).
\bibitem{mcphed}   S. K. Chin, N. A. Nicorovici, and R. C. McPhedran
Phys. Rev. E 49, 4590-4602 (1994)
\bibitem{mcphed2} N. A. Nicorovici, R. C. McPhedran, and R. Petit
Phys. Rev. E 49, 4563-4577 (1994)
\bibitem{mcphed3} N. A. Nicorovici, R. C. McPhedran, and L. C. Botten
Phys. Rev. E 52, 1135-1145 (1995)
\bibitem{moroz} A. Moroz, Optics Letters, Volume 26, Issue 15, 1119-1121
\bibitem{pendry} J. B. Pendry, A. J. Holden, W. J. Stewart and I. Youngs, Phys. Rev. Lett. {\bf 76}, 4773 (1996).
\bibitem{pendry2} Pendry J B, Holden A J, Robins D J and Stewart W J, 
IEEE Trans. Microw. Theory Tech. {\bf 47}, 2075 (1999).
\bibitem{moi1}  D. Felbacq, G. Bouchitt\'{e}, Waves in Random Media  {\bf 7}, 245 (1997).
\bibitem{moi2}  D. Felbacq, J. Phys. A {\bf 33}, 825 (2000).
\bibitem{moi3} D. Felbacq, J. Math. Phys. {\bf 43}, 52  (2002).
\bibitem{moi4}  D. Felbacq, G. Bouchitt\'{e}, Opt. Lett. {\bf 30}, 1189 (2005).
\bibitem{soukou} P. Marko, C. M. Soukoulis, Opt. Lett. {\bf 28},  846  (2003).
\bibitem{pokro} A. L. Pokrovsky and A. L. Efros, Phys. Rev. Lett. {\bf 89}, 093901 (2002).
\bibitem{smithtera} T. J. Yen et al., Science {\bf 303}, 1494 (2004).
\bibitem{obrien} S. O'Brien and J. B. Pendry, J. Phys.: Condens. Matter{\bf14}, 14035 (2002).
\bibitem{smith}R. A. Shelby, D. R. Smith, S. Schultz, Science {\bf 292}, 77 (2001).
\bibitem{val}P. M. Valanju, R. M. Walser, and A. P. Valanju, Phys. Rev. Lett. {\bf 88}, 187401 (2002).
\bibitem{muneg} M. Shamonin, E. Shamonina, V. Kalinin, and L. Solymar, J. Appl. Phys. 95, 3778 (2004). 
\bibitem{felbou} D. Felbacq, G. Bouchitté, Phys. Rev. Lett. {\bf 94}, 183902 (2005).
\bibitem{veselago} V. G. Veselago, Sov. Phys. Usp. {\bf 10}, 509 (1968). 
\bibitem{feljos}  D. Felbacq et al., J. Opt. Soc. Am. A  {\bf 11}, 2526 (1994). 
\end{thebibliography}
\end{document}